\documentclass[prb,preprintnumbers,amsmath,amssymb,floatfix]{revtex4}

\usepackage{graphicx}
\usepackage{subfigure}
\usepackage{dcolumn}

\begin{document}

\title{The quantum theory and topological features of photon}
\author{Xiang-Yao Wu$^{a}$\footnote{E-mail: xiangyaowu2006@163.com}, Ben-Shan Wu$^{a}$, Qing Pan$^{a}$ \\ Xiao-Jing Liu$^{a}$, Ji-Ping Liu$^{a}$, Xiao-Ru
Zhang$^{a}$ and Ji Ma$^{a}$}
 \affiliation{a. Institute of Physics, Jilin Normal
University, Siping 136000 China }

%%%%%%%
\begin{abstract}
In this paper, we have proposed the spinor wave equation of free
and non-free photon. On this basis, we given the spin operators
and spin wave functions of photon, and calculated the wave
function of photon in vacuum and medium. In addition, we have
given the quantum Berry phase and Chern number with the photon
wave function, which can be used to studied the quantum
topological features of photon in one-dimensional period medium.

\vskip 5pt



\end{abstract}

\maketitle

\vskip 8pt
 {\bf 1. Introduction} \vskip 8pt

The Dirac equation [1] is the important equation for the
relativistic particles of spin $\frac{1}{2}$. Within quantum
electrodynamics, The theoretical calculations and experimental
results are highly consistent. In addition, the Dirac equation is
the basis of the theory of electroweak interactions, quantum
chromodynamics and quantum hadrodynamics. The Klein-Gordon
equation [2, 3] is the fundamental equations of quantum field
theories, describes bosons with spin zero. The present work wants
to show that there exists a modification of the Klein-Gordon
equation, which includes the relativistic spin effects [4].

The Dirac equation is a first order differential equation, whereas
the Maxwell equations and Klein-Gordon equation correspond to a
second order differential equation for the photon field. After
Dirac discovered the relativistic equation for a particle with
spin $1/2$, much work was done to study spinor and vectors within
the Lorentz group theory for any spin particle

The initial motivation of this paper was the question whether one
can find a first order differential equation of photon. The
spinors are considered as fundamental physical quantity in quantum
field theory [5£¬6]. For this reason, we have proposed the spinor
wave equations of photon. In recent years, various novel
topological phenomena addressed in the condensed matter
physics[7-9] have been achieved in photonic systems, where
different types of topological insulators have all found their
counterparts[10]. The realization of classical analogues of
topological insulators in artificial crystals has been an emerging
research area [11-13]. Photonic topological insulators (PTIs) have
been theoretically proposed and experimentally demonstrated in
different photonic systems [14-16]. The Berry phase provides a
universal framework which relates the robust quantization of
physical observables at the boundaries of a non-interacting system
to the topological properties of the bulk.

\vskip 8pt
 {\bf 2. Relativistic spinor wave equation of free photon} \vskip 8pt

As is known to all, Dirac equation describes the particle of spin $\frac{1}%
{2}$ by factorizing Einstein's dispersion relation, such that the field
equation becomes the first order in time derivative [28]. Namely, Dirac
factorized the relativistic dispersion relation employing four by four
matrices, which is expressed as%

\begin{equation}
E^{2}-c^{2}\vec{p}\hspace{0.05in}^{2}-m_{0}^{2}c^{4}=(E-c\vec{p}\cdot
\vec{\alpha}-m_{0}c^{2}\beta)(E+c\vec{p}\cdot \vec{\alpha}+m_{0}c^{2}\beta)=0,
\end{equation}
thus we get
\begin{equation}
E-c\vec{\alpha}\cdot \vec{p}-m_{0}c^{2}\beta=0.
\end{equation}
By canonical quantization Eq. (2), i.e., $E\rightarrow i\hbar
\frac{\partial}{\partial t}$, $\vec{p}\rightarrow-i\hbar \nabla$,
we can obtain the Dirac spinor wave equation
\begin{equation}
i\hbar \frac{\partial}{\partial t}\psi(\vec{r}, t)=(-ic\hbar
\vec{\alpha}\cdot
\vec{\bigtriangledown}+m_{0}c^{2}\beta)\psi(\vec{r}, t),
\end{equation}
where $\vec{\alpha}$ and $\beta$ are Dirac matrices.

With Dirac's factorization approach, we can obtain the spinor wave
equation of free photon. For a photon, its mass $m_{0}=0$, Eq. (2)
becomes
\begin{equation}
E-c\vec{\alpha}\cdot \vec{p}=0.
\end{equation}
By canonical quantization Eq. (4), we obtain the spinor wave
equation of photon
\begin{equation}
i\hbar \frac{\partial}{\partial t}\psi(\vec{r}, t)=-ic\hbar
\vec{\alpha}\cdot \vec {\bigtriangledown}\psi(\vec{r},
t)=H\psi(\vec{r}, t),
\end{equation}
where $H=-ic\hbar \vec{\alpha}\cdot \vec{\bigtriangledown}$ is Hamiltonian
operator and $\psi$ is the spinor wave function of photon. For the proper
Lorentz group $L_{p}$, the irreducibility representations of spin $s=1$ photon
are $D^{10}$, $D^{01}$ and $D^{\frac{1}{2}\frac{1}{2}}$, respectively, and the
dimension of irreducibility representations corresponds to three, three and
four, respectively. We choose photon's spinor wave function as the basis
vector of three dimension irreducibility representation, i.e.
\begin{equation}
\psi(\vec{r}, t)=\left(
\begin{array}
[c]{c}%
\psi_{1}(\vec{r}, t)\\
\psi_{2}(\vec{r}, t)\\
\psi_{3}(\vec{r}, t)%
\end{array}
\right)  ,
\end{equation}
and the $\vec{\alpha}$ matrices are denoted by
\begin{equation}
\alpha_{x}= \left(
\begin{array}
[c]{ccc}%
0 & 0 & 0\\
0 & 0 & -i\\
0 & i & 0
\end{array}
\right)  ,\alpha_{y}= \left(
\begin{array}
[c]{ccc}%
0 & 0 & i\\
0 & 0 & 0\\
-i & 0 & 0
\end{array}
\right)  ,\alpha_{z}= \left(
\begin{array}
[c]{ccc}%
0 & -i & 0\\
i & 0 & 0\\
0 & 0 & 0
\end{array}
\right)  ,
\end{equation}
which are Hermitean matrices, $\vec{\alpha}^{\dag}=\vec{\alpha}$. The photon's
Hamiltonian operator is also Hermitean ${H}^{\dag}={H}.$

\vskip 8pt
 {\bf 3. The spin operators of photon} \vskip 8pt

In this section, we shall prove that the selection of
$\vec{\alpha}$ matrices in Eq. (7) is reasonable, and the Eqs.
(5), (6) and (7) are the spinor wave equation of free photon,
i.e., they are corresponding to the spinor wave equation of spin
$s=1$ and mass $m_{0}=0$.\newline

The equation (5) can be written as
\begin{equation}
i\hbar \frac{\partial}{\partial t}\psi(\vec{r}, t)=c(\vec{p}\cdot
\vec{\alpha})=H\psi(\vec{r}, t),
\end{equation}
where ${H}=c\vec{p}\cdot \vec{\alpha}$. The orbital angular momentum of photon
satisfies
\begin{align}
\frac{d}{dt}{L_{x}}  &  =\frac{1}{i\hbar}[{L_{x}},{H}]\nonumber \\
&  =c(\alpha_{y}{p_{z}}-\alpha_{z}{p_{y}})\nonumber \\
&  =c(\vec{\alpha}\times \vec{p})_{x},
\end{align}
so
\begin{equation}
\lbrack \vec{L},{H}]=i\hbar c(\vec{\alpha}\times \vec{p}).
\end{equation}
The Eq. (10) is shown that the orbital angular momentum of photon
is not conservation, but the total angular momentum of photon
should be conservative. Thus, photon should have an intrinsic
angular momentum, i.e., spin angular momentum $\vec{s}$, and the
total angular momentum of photon $\vec{J}$ is
\begin{equation}
{\vec{J}}={\vec{L}}+{\vec{s}},
\end{equation}
and $\vec{J}$ should be conservative $[{\vec{J}},{H}]=0$. Again
considering Eq. (10), we have
\begin{equation}
\lbrack \vec{s},{H}]=-[{\vec{L}},{H}]=-i\hbar c(\vec{\alpha}\times \vec{p}),
\end{equation}
where the spin component ${s}_{x}$ satisfies
\begin{align}
\lbrack{s}_{x},{H}]  &  =[{s}_{x},c\vec{\alpha}\cdot \vec{p}]=-i\hbar
c(\vec{\alpha}\times \vec{p})_{x}\nonumber \\
&  =i\hbar c(\alpha_{z}{p}_{y}-\alpha_{y}{p}_{z}),
\end{align}
i.e.,
\begin{align}
\lbrack{s}_{x},c\alpha_{x}p_{x}+c\alpha_{y}p_{y}+c\alpha_{z}p_{z}]  &
=c[s_{x},\alpha_{x}]p_{x}+c[s_{x},\alpha_{y}]p_{y}+c[s_{x},\alpha_{z}%
]p_{z}\nonumber \\
&  =i\hbar c(\alpha_{z}{p}_{y}-\alpha_{y}{p}_{z}).
\end{align}
Comparing with the both sides of Eq. (14), we obtain the
commutation relations
\begin{equation}
\lbrack s_{x},\alpha_{x}]=0,[s_{x},\alpha_{y}]=i\hbar \alpha_{z},[s_{x}%
,\alpha_{z}]=-i\hbar \alpha_{y}.
\end{equation}
Similarly, it is obtained
\begin{equation}
\lbrack s_{y},\alpha_{y}]=0,[s_{y},\alpha_{x}]=-i\hbar \alpha_{z},[s_{y}%
,\alpha_{z}]=i\hbar \alpha_{x},
\end{equation}
and
\begin{equation}
\lbrack s_{z},\alpha_{z}]=0,[s_{z},\alpha_{x}]=i\hbar \alpha_{y},[s_{z}%
,\alpha_{y}]=-i\hbar \alpha_{x}.
\end{equation}
According to commutation relations in Eqs. (15), (16) and (17) and
using Eq. (7), we can calculate the spin matrices $\vec {s}$ of
photon, they are
\begin{equation}
s_{x}=\left(
\begin{array}
[c]{ccc}%
a & 0 & 0\\
0 & a & -i\hbar \\
0 & i\hbar & a
\end{array}
\right)  ,s_{y}=\left(
\begin{array}
[c]{ccc}%
b & 0 & i\hbar \\
0 & b & 0\\
-i\hbar & 0 & b
\end{array}
\right)  ,s_{z}=\left(
\begin{array}
[c]{ccc}%
c & -i\hbar & 0\\
i\hbar & c & 0\\
0 & 0 & c
\end{array}
\right)  ,
\end{equation}
where $a$, $b$ and $c$ are to be determined by their eigenequations.

Using the photon spin matrices $s_{x}$, $s_{y}$, $s_{z}$, their eigenvalues
should be $\pm \hbar$. For the $s_{x}$ eigenvalue problem, we have
\begin{equation}
\left(
\begin{array}
[c]{ccc}%
a & 0 & 0\\
0 & a & -i\hbar \\
0 & i\hbar & a
\end{array}
\right)  \left(
\begin{array}
[c]{c}%
\psi_{1}\\
\psi_{2}\\
\psi_{3}%
\end{array}
\right)  =\lambda_{1}\left(
\begin{array}
[c]{c}%
\psi_{1}\\
\psi_{2}\\
\psi_{3}%
\end{array}
\right)  ,
\end{equation}
and its characteristic equation is
\begin{equation}
\left \vert
\begin{array}
[c]{ccc}%
a-\lambda_{1} & 0 & 0\\
0 & a-\lambda_{1} & -i\hbar \\
0 & i\hbar & a-\lambda_{1}%
\end{array}
\right \vert =0,
\end{equation}
i.e.,
\begin{equation}
(a-\lambda_{1})[(a-\lambda_{1})^{2}-\hbar^{2}]=0. \label{30}%
\end{equation}
In order to get the eigenvalues $\lambda_{1}=\pm \hbar$, we should
set $a=0.$ In the similar method, we also have $b=0$ and $c=0$.
Finally, we may obtain the spin matrices of photon in Eq. (18) and
after calculation these matrices square are
\begin{equation}
{\vec{s}}\hspace{0.05in}^{2}=s_{x}^{2}+s_{y}^{2}+s_{z}^{2}=2\left(
\begin{array}
[c]{ccc}%
1 & 0 & 0\\
0 & 1 & 0\\
0 & 0 & 1
\end{array}
\right)  \hbar^{2}=s(s+1)\hbar^{2}\left(
\begin{array}
[c]{ccc}%
1 & 0 & 0\\
0 & 1 & 0\\
0 & 0 & 1
\end{array}
\right)  ,
\end{equation}
i.e., $s=1$, the spin matrices Eq. (18) with $a=0,$ $b=0$ and
$c=0$ are photon's spin matrices. So Eqs. (5), (6) and (7) are
just the spinor wave equation of free photon. According to Eq.
(7), we find
\begin{equation}
s_{x}=\hbar\alpha_{x},\hspace{0.2in}s_{y}=\hbar\alpha_{y},\hspace{0.2in}s_{z}=\hbar\alpha
_{z}.
\end{equation}

\vskip 8pt

\vskip 8pt
 {\bf 5. The probability conservation equation of photon} \vskip 8pt

In the following section, we shall give the probability density and
probability conservation equation of photon.\newline

The hermitian conjugate of Eq. (5) is
\begin{equation}
-i\hbar \frac{\partial \psi^{+}}{\partial t}=i\hbar c\vec{\nabla}\psi^{+}%
\cdot \vec{\alpha},
\end{equation}
right multiplying Eq. (24) by $\psi$
\begin{equation}
-i\hbar \frac{\partial \psi^{+}}{\partial t}\psi=i\hbar
c\vec{\nabla}\psi ^{+}\cdot \vec{\alpha}\psi,
\end{equation}
and left multiplying Eq. (5) by $\psi^{+}$
\begin{equation}
i\hbar \psi^{+}(\frac{\partial \psi}{\partial t})=-i\hbar
c\psi^{+}\vec{\alpha }\cdot \vec{\nabla}\psi,
\end{equation}
we get
\begin{equation}
i\hbar(\psi^{+}\frac{\partial \psi}{\partial t}+\frac{\partial \psi^{+}%
}{\partial t}\psi)+i\hbar
c\psi^{+}\vec{\alpha}\cdot(\vec{\nabla}\psi)+i\hbar
c\vec{\nabla}\psi^{+}\cdot \vec{\alpha}\psi=0,
\end{equation}
or
\begin{equation}
\frac{1}{c}\frac{\partial}{\partial t}(\psi^{+}\psi)+\psi^{+}\vec{\alpha}%
\cdot(\vec{\nabla}\psi)+(\vec{\nabla}\psi^{+})\cdot \vec{\alpha}\psi=0,
\end{equation}
and obtain the probability conservation equation of photon
\begin{equation}
\frac{\partial \rho}{\partial t}+\nabla \cdot \vec{J}=0.
\end{equation}
Where $\rho=\psi^{+}\psi$ and $\vec{J}=c\psi^{+}\vec{\alpha}\psi$
are the probability and probability current density of photon,
respectively, which are expressed by the spinor wave functions
$\psi$ and $\psi^{+}$ of photon, and the photon probability
density $\rho \geq 0$.

When photon is incident to a uniform medium of finite volume,
there are incident, reflection and transmission photon in the
medium, we can define the quantum transmission coefficient $t$ and
reflection coefficient $r$, they are
\begin{equation}
t=|\frac{J_{Dx}}{J_{Ix}}|=|\frac{\psi^{+}_{D}(x)\alpha_{x}\psi_{D}(x)}{\psi^{+}_{I}(x)\alpha_{x}\psi_{I}(x)}|,
\end{equation}
and
\begin{equation}
r=|\frac{J_{Rx}}{J_{Ix}}|=|\frac{\psi^{+}_{R}(x)\alpha_{x}\psi_{R}(x)}{\psi^{+}_{I}(x)\alpha_{x}\psi_{I}(x)}|.
\end{equation}
Where $\psi^{+}_{I}(x)$, $\psi^{+}_{R}(x)$ and $\psi^{+}_{D}(x)$
are the incident, reflection and transmission wave functions of
photon in the medium.

\vskip 8pt
 {\bf 6. The plane wave solution and helicity of free photon} \vskip 8pt

For the spin $\frac{1}{2}$ Dirac particle, there are the plane
wave solutions corresponding to positive energy and negative
energy. For the photon, there are also the plane wave solutions
corresponding to positive energy and negative energy. Based on the
above discussion, we have the spinor equation of free photon
\begin{equation}
i\hbar \frac{\partial}{\partial t}\psi=H\psi,
\end{equation}
where
\begin{equation}
H=c\hspace{0.02in}\vec{{\alpha}}\cdot
\vec{p},\hspace{0.1in}\psi=\left(
\begin{array}
[c]{c}%
\psi_{1}\\
\psi_{2}\\
\psi_{3}%
\end{array}
\right)  .
\end{equation}
Since $\frac{\partial H}{\partial t}=0$ and $[\vec{p},H]=0$, the
photon energy $E$ and momentum $\vec{p}$ are conserved quantity,
they have common eigenstate, namely,
\begin{equation}
\psi_{E,\vec{p}}(\vec{r},t)=u(\vec{p})e^{i(\vec{p}\cdot
\vec{r}-Et)/\hbar},
\label{59}%
\end{equation}
where
\begin{equation}
u(\vec{p})=\left(
\begin{array}
[c]{c}%
u_{1}(\vec{p})\\
u_{2}(\vec{p})\\
u_{3}(\vec{p})
\end{array}
\right)  .
\end{equation}
Substituting Eqs. (34) and (35) into (32) yields
\begin{equation}
c\hspace{0.02in}\vec{{\alpha}}\cdot \vec{p}\hspace{0.1in}u(\vec{p}%
)=E\hspace{0.02in}u(\vec{p}),
\end{equation}
i.e.,
\begin{equation}
\left(
\begin{array}
[c]{ccc}%
0 & -ic\hspace{0.01in}p_{z} & ic\hspace{0.01in}p_{y}\\
ic\hspace{0.01in}p_{z} & 0 & -ic\hspace{0.01in}p_{x}\\
-ic\hspace{0.01in}p_{y} & ic\hspace{0.01in}p_{x} & 0
\end{array}
\right)  \left(
\begin{array}
[c]{c}%
u_{1}\\
u_{2}\\
u_{3}%
\end{array}
\right)  =E\left(
\begin{array}
[c]{c}%
u_{1}\\
u_{2}\\
u_{3}%
\end{array}
\right)  .
\end{equation}
Expanding Eq. (37), we get
\begin{align}
Eu_{1}+icp_{z}u_{2}-icp_{y}u_{3}  &  =0,\label{63}\\
icp_{z}u_{1}-Eu_{2}-icp_{x}u_{3}  &  =0,\label{64}\\
icp_{y}u_{1}-icp_{x}u_{2}+Eu_{3}  &  =0, \label{65}%
\end{align}
by the sufficient necessary condition of nonzero solution of
$u_{1}$, $u_{2}$ and $u_{3}$, we get
\begin{equation}
\left \vert
\begin{array}
[c]{ccc}%
E & icp_{z} & -icp_{y}\\
icp_{z} & -E & -icp_{x}\\
\hspace{0.05in}icp_{y}\hspace{0.1in} & -icp_{x} & E
\end{array}
\right \vert =0,
\end{equation}
with the eigenvalues $E$ being
\begin{equation}
E_{1}=+c|\vec{p}|,\hspace{0.2in}E_{2}=-c|\vec{p}|.
\end{equation}
From Eqs. (39) and (40), we have
\begin{equation}
icp_{z}p_{y}u_{1}-Ep_{y}u_{2}-icp_{x}p_{y}u_{3}=0,
\end{equation}
\begin{equation}
icp_{y}p_{z}u_{1}-icp_{x}p_{z}u_{2}+Ep_{z}u_{3}=0,
\end{equation}
taking the difference of Eqs. (43) and (44), it is obtained that
\begin{equation}
(Ep_{y}-icp_{x}p_{z})u_{2}+(Ep_{z}+icp_{x}p_{y})u_{3}=0,
\end{equation}
or
\begin{equation}
\frac{u_{2}}{u_{3}}=-\frac{Ep_{z}+icp_{x}p_{y}}{Ep_{y}-icp_{x}p_{z}}.
\end{equation}
Substituting Eq. (47) into Eq. (38), there is
\begin{equation}
\frac{u_{1}}{u_{3}}=\frac{ic(p_{y}^{2}+p_{z}^{2})}{Ep_{y}-icp_{x}p_{z}}.
\end{equation}
According to Eqs. (46) and (47), there is
\begin{equation}
\frac{u_{1}}{u_{2}}=-\frac{ic(p_{y}^{2}+p_{z}^{2})}{Ep_{z}+icp_{x}p_{y}},
\end{equation}
thus the $u(p)$ can be written as
\begin{equation}
u(p)=N\left(
\begin{array}
[c]{c}%
ic(p_{y}^{2}+p_{z}^{2})\\
-(Ep_{z}+icp_{x}p_{y})\\
Ep_{y}-icp_{x}p_{z}%
\end{array}
\right)  ,
\end{equation}
where $N$ is normalization constant. Using $u^{+}(p)u(p)=1$, we
can obtain
\begin{equation}
2E^{2}N^{2}(p_{y}^{2}+p_{z}^{2})=1,
\end{equation}
and the normalization constant is
\begin{equation}
N=\sqrt{\frac{1}{2E^{2}(p_{y}^{2}+p_{z}^{2})}}.
\end{equation}
Therefore,
\begin{equation}
u(p)=\sqrt{\frac{1}{2E^{2}(p_{y}^{2}+p_{z}^{2})}}\left(
\begin{array}
[c]{c}%
ic(p_{y}^{2}+p_{z}^{2})\\
-(Ep_{z}+icp_{x}p_{y})\\
Ep_{y}-icp_{x}p_{z}%
\end{array}
\right)  ,
\end{equation}
and
\begin{equation}
\psi_{E,\vec{P}}(\vec{r},t)=\sqrt{\frac{1}{2E^{2}(p_{y}^{2}+p_{z}^{2})}%
}\left(
\begin{array}
[c]{c}%
ic(p_{y}^{2}+p_{z}^{2})\\
-(Ep_{z}+icp_{x}p_{y})\\
Ep_{y}-icp_{x}p_{z}%
\end{array}
\right)  e^{i(\vec{p}\cdot \vec{r}-Et)/\hbar},
\end{equation}
which is just the plane wave solution of free photon. Substituting
$E_{1}=+c|\vec{p}|$ and $E_{2}=-c|\vec{p}|$ into Eq. (53), we can
obtain the photon plane wave solutions of positive energy and
negative energy.

The polarization vector acts for the photon as the "spin part" of
the wave function, the polarization of the photon is in a certain
relationship to photon's helicity. For a given momentum photon, it
has two different polarizations, which may be taken to be two
mutually perpendicular linear polarizations, and the two circular
polarizations having opposite directions of rotation, i.e., the
right-hand and left-hand circular polarizations. By the helicity
of photon, we can obtain the two different polarizations of
photon. The helicity is defined as the projection of spin in the
momentum direction, i.e.
\begin{equation}
h=\frac{\vec{\alpha}\cdot \vec{p}} {|\vec{p}|},
\end{equation}
with Eq. (33), the helicity becomes
\begin{equation}
h=\frac{H} {c|\vec{p}|}.
\end{equation}
Since the eigenvalues of $H$ are $+c|\vec{p}|$ and $-c|\vec{p}|$,
the eigenvalues of $h$ are $+1$ and $-1$, which corresponds to a
right-handed and left-handed, transverse, circularly polarized
helicity state of photon.

\vskip 8pt
 {\bf 6. The spin wave functions of photon} \vskip 8pt

In the section 3, we have given the spin operator of photon, we
can calculate the spin wave functions of photon. From Eqs. (18)
and (22), we find that $\vec{s}\hspace{0.05in}^{2}$ commutes with
$s_{x}$, $s_{y}$ and $s_{z}$,
respectively. Thus we can calculate the common eigenstates of $\vec{s}%
\hspace{0.05in}^{2}$ and $s_{z}$, expressed as
\begin{equation}
\vec{s}\hspace{0.05in}^{2}\chi_{\mu}=2\hbar^{2}\chi_{\mu},
\end{equation}
\begin{equation}
s_{z}\chi_{\mu}=\mu \hbar \chi_{\mu},
\end{equation}
where $(\chi_{\mu})^{T}=(\varphi_{1},\varphi_{2},\varphi_{3})$ is
their common eigenstate. Considering Eq. (38), we easily rewritten
as
\begin{equation}
\left(
\begin{array}
[c]{ccc}%
0 & -i\hbar & 0\\
i\hbar & 0 & 0\\
0 & 0 & 0
\end{array}
\right)  \left(
\begin{array}
[c]{c}%
\varphi_{1}\\
\varphi_{2}\\
\varphi_{3}%
\end{array}
\right)  =\mu\hbar \left(
\begin{array}
[c]{c}%
\varphi_{1}\\
\varphi_{2}\\
\varphi_{3}%
\end{array}
\right)  ,
\end{equation}
and according to its characteristic equation, the eigenvalues $\mu$ are
\begin{equation}
\mu_{1}=0,\hspace{0.1in}\mu_{2}=1,\hspace{0.1in}\mu_{3}=-1,
\end{equation}
Substituting $\mu_{1}=0$ into Eq. (58), we get
\begin{equation}
\left \{
\begin{array}
[c]{ll}%
-i\varphi_{2}=0 & \\
\hspace{0.1in}i\varphi_{1}=0 &
\end{array}
\right.  ,
\end{equation}
i.e.,
\begin{equation}
\varphi_{1}=\varphi_{2}=0,\varphi_{3}\neq0,
\end{equation}
and the normalization spin wave function is
\begin{equation}
\chi_{0}=\left(
\begin{array}
[c]{c}%
0\\
0\\
1
\end{array}
\right)  .
\end{equation}
Similarly, by substituting $\mu_{2}=1$ and $\mu_{3}=-1$ into Eq.
(58), respectivley, the corresponding normalization spin wave
function is
\begin{equation}
\chi_{1}=-\frac{1}{\sqrt{2}}\left(
\begin{array}
[c]{c}%
1\\
i\\
0
\end{array}
\right)  , \label{91}%
\end{equation}
and
\begin{equation}
\chi_{-1}=\frac{1}{\sqrt{2}}\left(
\begin{array}
[c]{c}%
1\\
-i\\
0
\end{array}
\right)  .
\end{equation}
These spin wave functions satisfy the normalization condition
\begin{equation}
\sum_{\alpha}\chi_{\mu}^{\ast}(\alpha)\chi_{\mu^{\prime}}(\alpha)=\delta
_{\mu \mu^{\prime}}.
\end{equation}
By the spin wave functions of single-photon, we can obtain the spin wave
functions of two-photon and multiphoton, and can further give the spin
entanglement states of two-photon and multiphoton, which can be used in
quantum information.

\vskip 8pt
 {\bf 7. The spinor wave equation of non-free photon} \vskip 8pt

In the sections above, we have given the spinor wave equation of free photon,
i.e., the photon is in the vacuum. When photon is in the medium, it becomes
non-free photon. Next, we shall give the spinor wave equation of non-free photon.

For the non-free particle, the Einstein's dispersion relation is
\begin{equation}
(E-V)^{2}=c^{2}\vec{p}\hspace{0.05in}^{2}+m_{0}^{2}c^{4}.
\end{equation}
Factorizing Eq. (66), we obtain
\begin{equation}
(E-V)^{2}-c^{2}\vec{p}\hspace{0.05in}^{2}-m_{0}^{2}c^{4}=(E-V-c\vec{p}%
\cdot \vec{\alpha}-m_{0}c^{2}\beta)(E-V+c\vec{p}\cdot \vec{\alpha}+m_{0}%
c^{2}\beta)=0.
\end{equation}
For a photon, $m_{0}=0$, Eq. (67) becomes
\begin{equation}
(E-V-c\vec{p}\cdot \vec{\alpha})(E-V+c\vec{p}\cdot
\vec{\alpha})=0,
\end{equation}
or
\begin{equation}
(E-V-c\vec{p}\cdot \vec{\alpha})=0.
\end{equation}
According to canonical quantization, Eq. (69) turns into
\begin{equation}
i\hbar \frac{\partial}{\partial t}\psi(\vec{r}, t)=-ic\hbar
\vec{\alpha}\cdot \vec{\nabla }\psi(\vec{r}, t)+V\psi(\vec{r}, t).
\end{equation}
Considering that the potential energy of photon in medium is [29]
\begin{equation}
V=\hbar \omega(1-n),
\end{equation}
where $n$ is the medium refractive index. Substituting Eq. (71)
into (70), the spiron equation of photon in medium is expressed as
\begin{equation}
i\hbar \frac{\partial}{\partial t}\psi(\vec{r}, t)=-ic\hbar
\vec{\alpha}\cdot \vec{\nabla }\psi(\vec{r}, t)+\hbar
\omega(1-n)\psi(\vec{r}, t).
\end{equation}
Using the method of separation variable
\begin{equation}
\psi(\vec{r},\vec{t})=\psi(\vec{r})f(t),
\end{equation}
Eq. (72) becomes
\begin{equation}
f(t)=f_{0}e^{-\frac{i}{\hbar}Et},
\end{equation}
\begin{equation}
\lbrack-ic\hbar \vec{\alpha}\cdot \vec{\nabla}+\hbar
\omega(1-n)]\psi(\vec {r})=E\psi(\vec{r}),
\end{equation}
where $E$ is the total energy of photon in medium, and $n$ is the
refractive index of medium. The Eqs. (72) and (75) are the spiron
wave equations of time-dependent and time-independent of photon in
the medium, which can be used to study the quantum property of
photon in medium.

 \vskip 8pt
 {\bf 8. The plane wave solution of photon in medium} \vskip 8pt

For the spin $\frac{1}{2}$ Dirac electron, there are the plane
wave solutions corresponding to positive energy and negative
energy. For the photon, there are also the plane wave solutions
corresponding to positive energy and negative energy. The Eq. (72)
Hamiltonian operator and spiron wave function are
\begin{equation}
H=c\hspace{0.02in}\vec{{\alpha}}\cdot \vec{p}+\hbar
\omega(1-n),\hspace{0.2in}\psi(\vec{r}, t)=\left(
\begin{array}
[c]{c}%
\psi_{1}(\vec{r}, t)\\
\psi_{2}(\vec{r}, t)\\
\psi_{3}(\vec{r}, t)%
\end{array}
\right),
\end{equation}
since $\frac{\partial H}{\partial t}=0$ and $[\vec{p},H]=0$, the
photon energy $E$ and momentum $\vec{p}$ are conserved quantity,
they have common eigenstate, namely,
\begin{equation}
\psi(\vec{r},t)=u(\vec{p})e^{i(\vec{p}\cdot
\vec{r}-Et)/\hbar}=\psi(\vec{r})e^{-i\omega
t}=u(\vec{k})e^{i(\vec{k}\cdot \vec{r}-\omega t)},
\end{equation}
where $\vec{k}=\vec{p}/{\hbar}$, $\omega={E}/{\hbar}$ and
\begin{equation}
\psi(\vec{r})=\left(
\begin{array}
[c]{c}%
\psi_{1}\\
\psi_{2}\\
\psi_{3}
\end{array}
\right)=\left(
\begin{array}
[c]{c}%
u_{1}\\
u_{2}\\
u_{3}
\end{array}
\right)e^{i\vec{k}\cdot \vec{r}}=u(\vec{k})e^{i\vec{k}\cdot
\vec{r}},
\end{equation}
substituting Eqs. (78) into (72) yields
\begin{eqnarray}
ic\hbar\alpha\cdot\bigtriangledown\psi(\vec{r})+(E-\hbar\omega(1-n))\psi(\vec{r})=0,
\end{eqnarray}
substituting Eqs. (7) and (78) into (79), there are
\begin{eqnarray}
(E-(1-n)\hbar\omega)u_1+ic\hbar{k_z}{u_2}-ic\hbar{k_y}{u_3}=0,
\end{eqnarray}
\begin{eqnarray}
-ic\hbar{k_z}{u_1}+(E-(1-n)\hbar\omega)u_2+ic\hbar{k_x}{u_3}=0,
\end{eqnarray}
and
\begin{eqnarray}
ic\hbar{k_y}{u_1}-ic\hbar{k_x}{u_2}+(E-(1-n)\hbar\omega)u_3=0,
\end{eqnarray}
the necessary and sufficient conditions for a non-zero solution of
Eqs. (80)-(82) is
\begin{equation}
\left \vert
\begin{array}
[c]{ccc}%
E-(1-n)\hbar\omega & ic\hbar{k_z} & -ic\hbar{k_y}\\
-ic\hbar{k_z} & E-(1-n)\hbar\omega & ic\hbar{k_x} \\
ic\hbar{k_y} & -ic\hbar{k_x} & E-(1-n)\hbar\omega%
\end{array}
\right \vert =0,
\end{equation}
spreading Eq. (83), we obtain
\begin{eqnarray}
E_1=\hbar\omega(1-n),
\end{eqnarray}
\begin{eqnarray}
E_2=\hbar\omega(1-2n),
\end{eqnarray}
and
\begin{eqnarray}
E_3=\hbar\omega,
\end{eqnarray}
since $E_1< 0$ and $E_2< 0$, the solution $E_3=\hbar\omega$ is
reasonable. Substituting $E_3$ into (80)-(82), we get
\begin{eqnarray}
\omega n u_1+ick_zu_2-ick_yu_3=0,
\end{eqnarray}
\begin{eqnarray}
ick_zu_1-\omega n u_2-ick_xu_3=0,
\end{eqnarray}
and
\begin{eqnarray}
ick_yu_1-ick_xu_2+\omega n u_3=0.
\end{eqnarray}
From Eq. (87) to (89), we obtain the ratio
\begin{eqnarray}
\frac{u_2}{u_3}=\frac{ick_xk_y+\omega n k_z}{ick_xk_z-\omega n
k_y},
\end{eqnarray}
and
\begin{eqnarray}
\frac{u_1}{u_2}=-\frac{ic(k^2_y+k^2_z)}{ick_xk_y+\omega n k_z},
\end{eqnarray}
and the $u(\vec{k})$ spinor is
\begin{equation}
\left(
\begin{array}
[c]{c}%
u_{1}\\
u_{2}\\
u_{3}%
\end{array}
\right)=A\left(
\begin{array}
[c]{c}%
-ic(k^2_y+k^2_z)\\
ick_xk_y+\omega n k_z\\
ick_xk_z-\omega n
k_y%
\end{array}
\right),
\end{equation}
its normalization form is
\begin{equation}
\left(
\begin{array}
[c]{c}%
u_{1}\\
u_{2}\\
u_{3}%
\end{array}
\right)=\frac{1}{ck\sqrt{2(k^2_y+k^2_z)}}\left(
\begin{array}
[c]{c}%
-ic(k^2_y+k^2_z)\\
ick_xk_y+\omega n k_z\\
ick_xk_z-\omega n
k_y%
\end{array}
\right).
\end{equation}
In the medium of refractive index $n$, when the photon propagates
in the $x-y$ plane, the wave vector
$\vec{k}=k_{x}\vec{i}+k_{y}\vec{j}$, where
$k_x=k\cos\theta=\frac{\omega}{c}n\cos\theta$,
$k_y=k\sin\theta=\frac{\omega}{c}n\sin\theta$ and $\theta$ is the
angle between $\vec{k}$ and $x$ axis, the Eq. (93) becomes
\begin{equation}
\left(
\begin{array}
[c]{c}%
u_{1}\\
u_{2}\\
u_{3}%
\end{array}
\right)=\frac{1}{\sqrt{2}\omega n\cdot k_y}\left(
\begin{array}
[c]{c}%
-ick^2_y\\
ick_xk_y\\
-\omega n k_y%
\end{array}
\right).
\end{equation}
In the vacuum, the medium refractive index $n=1$, the Eq. (94)
becomes
\begin{equation}
\left(
\begin{array}
[c]{c}%
u_{1}\\
u_{2}\\
u_{3}%
\end{array}
\right)=\frac{1}{\sqrt{2}\omega}\left(
\begin{array}
[c]{c}%
-ick_y\\
ick_x\\
-\omega %
\end{array}
\right)=\frac{1}{\sqrt{2}}\left(
\begin{array}
[c]{c}%
-i\sin\theta\\
i\cos\theta\\
-1 %
\end{array}
\right),
\end{equation}
the plane wave solution of photon in the vacuum or uniform medium
is
\begin{equation}
\psi(\vec{r},t)=\left(
\begin{array}
[c]{c}%
u_{1}\\
u_{2}\\
u_{3}%
\end{array}
\right)  e^{i(\vec{k}\cdot \vec{r}-\omega t)}.
\end{equation}

\vskip 8pt
 {\bf 9. The Lagrangean density of photon spinor equation} \vskip 8pt

Before giving the second quantization of the field of a photon
particle, we must deepen our understanding of the formal
properties of the spinor equation of photon. We start by casting
the formalism of the spinor equation of photon into Hamiltonian
form, because this is necessary for canonical quantization. For
the photon spinor equation (4), we may give its Lagrangean
density. In the natural unit, $\hbar=c=1$, the Eq. (5) can be
written as
\begin{equation}
(\partial_{t}+\vec{\alpha}\cdot \vec{\nabla})\psi=0,
\end{equation}
or
\begin{equation}
\beta^{\mu}\partial_{\mu}\psi=0,
\end{equation}
where $\beta^{\mu}=(I,\vec{\alpha})$ and $I$ is a $3\times3$ unit
matrix. The adjoint of Eq. (98) is
\begin{equation}
\partial_{t}\psi^{+}+\vec{\nabla}\psi^{+}\cdot \vec{\alpha}=0,
\end{equation}
i.e.,
\begin{equation}
\partial_{\mu}\psi^{+}\beta^{\mu}=0,
\end{equation}
the Lagrangean density of photon can be taken as
\begin{equation}
{\mathcal{L}}=\psi^{+}\beta^{\mu}\partial_{\mu}\psi,
\end{equation}
Using Eq. (101), we have
\begin{equation}
\frac{\partial{\mathcal{L}}}{\partial \psi}=0,\hspace{0.1in}\frac
{\partial{\mathcal{L}}}{\partial \partial_{\mu}\psi}=\psi^{+}\beta^{\mu},
\end{equation}
and
\begin{equation}
\frac{\partial{\mathcal{L}}}{\partial \psi^{+}}=\beta^{\mu}\partial_{\mu}%
\psi,\hspace{0.1in}\frac{\partial{\mathcal{L}}}{\partial
\partial_{\mu}\psi ^{+}}=0.
\end{equation}
Substituting Eqs. (102) and (103) into Lagrangean equations
\begin{equation}
\frac{\partial{\mathcal{L}}}{\partial
\psi}-\partial_{\mu}\frac{\partial {\mathcal{L}}}{\partial
\partial_{\mu}\psi}=0,
\end{equation}
and
\begin{equation}
\frac{\partial{\mathcal{L}}}{\partial
\psi^{+}}-\partial_{\mu}\frac {\partial{\mathcal{L}}}{\partial
\partial_{\mu}\psi^{+}}=0,
\end{equation}
we can obtain the spinor photon wave equation (98) and its adjoint
equation (100).

To go over to a Hamiltonian formalism, the momentum canonically conjugate to
$\psi$ is
\begin{equation}
\pi=\frac{\partial{\mathcal{L}}}{\partial \dot{\psi}}=\psi^{+},
\end{equation}
the hamiltonian density is
\begin{equation}
\mathcal{H}=\pi \dot{\psi}-\mathcal{L}=-\psi^{+}\vec{\alpha}\cdot
\vec{\nabla }\psi.
\end{equation}
The Hamiltonian (101) is suitable for the quantization of photon
field.

\vskip 12pt
 {\bf 10. The topological features of photon in periodic potential} \vskip 8pt
The study of topological properties in condensed matter physics
emerges from the discovery of quantum hall effect and quantum
anomalous hall effect, which have attracted great attentions in
recent years [17, 18]. Recent studies have discovered that these
topological properties exist also in photonic systems [19, 20] and
acoustic systems [21]. In photonic systems, the Zak phase and
Chern number can also be obtained from the photonic Bloch state of
one-dimensional photonic crystal [22], which are calculated by
classic electrical field, they are the classical Zak phase and
Chern number. In the following, we shall study the quantum Zak
phase and Chern number when a photon in one-dimensional photonic
crystal with the photon wave function. The photon quantum wave
equation is
\begin{equation}
i\hbar \frac{\partial}{\partial t}\psi(x, t)=(-ic\hbar
\vec{\alpha}\cdot \vec{\nabla }+V(x))\psi(x, t),
\end{equation}
with the method of separation variable
\begin{equation}
\psi(x,\vec{t})=\psi(x)f(t),
\end{equation}
Eq. (108) becomes
\begin{equation}
\lbrack-ic\hbar \vec{\alpha}\cdot
\vec{\nabla}+V(x)]\psi_{m}(x)=E_{m}\psi_{m}(x),
\end{equation}
where $m$ labels energy band and $E_{m}$ is the energy spectrum
for the band $m$. The Eq. (110) describes the quantum peculiarity
of Bloch photon in period potential field $V(x)$, it has the
solution of Bloch wave, it is
\begin{equation}
\psi_{m}(x)=e^{ikx}u_{mk}(x),
\end{equation}
where $k$ is Bloch vector, and $u_{mk}(x)$ is the period function,
i.e.,
\begin{equation}
u_{mk}(x+na)=u_{mk}(x),
\end{equation}
where $a$ is the period and $n$ is a integer. In the medium of
refractive index $n(x)$, the photon potential energy is
\begin{equation}
V(x)=\hbar \omega(1-n(x)).
\end{equation}
For media $A$ and $B$, which refractive index are $n_a$, $n_b$ and
their thickness are $a$, $b$, respectively are constituted
one-dimensional photonic crystals, the photon potential energy in
one-dimensional photonic crystals is the period potential, it is
\begin{equation}
V(x+(a+b))=V(x).
\end{equation}
The quantum Berry phase is defined as
\begin{equation}
\gamma_{m}=\int_{-\pi/(a+b)}^{\pi/(a+b)}X_{m}(k)dk,
\end{equation}
where the quantity $X_{m}(k)$ is
\begin{equation}
X_{m}(k)=\frac{2\pi}{a+b}\int_{0}^{a+b}u_{mk}^{*}(x)i\frac{\partial
u_{mk}(x)}{\partial k}dx.
\end{equation}
The quantum Chern number on a surface enclosing the point is

\begin{equation}
C_{m}=\frac{1}{2\pi}\oint_{S}\mathbf{F}_{m}(\mathbf{k})\cdot
d\mathbf{S}
\end{equation}

Here $S$ denotes the integration surface, and
$\mathbf{F}_{m}(\mathbf{k})=\nabla\times\langle u_{mk}(x)\mid
i\partial_{k} \mid u_{mk}(x) \rangle$ is the Berry curvature of
the n-th band with the wave function $u_{mk}(x)$, where the band
indices n for the lower, middle, and higher bands are denoted as
$-$, $0$, and $+$, respectively.

From Eqs. (112) to (116), we can calculate the quantum Zak phase
of photon in the period potential field, and we can calculate the
quantum Chern number with Eq. (117).

 \vskip 8pt
 {\bf 11. Conclusions} \vskip 8pt

In this paper, we have proposed the spinor wave equation of free
photon, and given the spin operators, spin wave functions and
space wave functions. We have calculated photon helicity and found
left-handed and right-handed photon. In addition, we have given
the Hamiltonian density of free photon, which is suitable for the
quantization of photon field. Otherwise£¬we have further given the
spinor wave equation of non-free photon, which should be used to
study the quantum property of photon in medium. Using the
single-photon spin wave function, we can study two-photon or
multiple-photon spin wave function, and further give the spin
entanglement states of multiple-photon, which should be applied in
quantum communication.

\newpage

 \vskip 8pt
 {\bf 11. Acknowledgment} \vskip 8pt

This work was supported by the Research innovation project of
Jilin Normal University (no.201627).

\end{document}